\documentclass{elsart41}
\usepackage{graphics}
\usepackage{graphicx}
\usepackage{amssymb}
\usepackage{times}
\newcommand{\nag}{\phantom{\dag}}
\begin{document}

\begin{frontmatter}

% Title, authors and addresses

% use the thanksref command within \title, \author or \address for footnotes;
% use the corauthref command within \author for corresponding author footnotes;
% use the ead command for the email address,
% and the form \ead[url] for the home page:
% \title{Title\thanksref{label1}}
% \thanks[label1]{}
% \author{Name\corauthref{cor1}\thanksref{label2}}
% \ead{email address}
% \ead[url]{home page}
% \thanks[label2]{}
% \corauth[cor1]{}
% \address{Address\thanksref{label3}}
% \thanks[label3]{}

\title{Optical response of many-polaron systems}

% use optional labels to link authors explicitly to addresses:
% \author[label1,label2]{}
% \address[label1]{}
% \address[label2]{}
\author[Er]{G. Wellein\corauthref{Name1}}, \ead{g.wellein@rrze.uni-erlangen.de}
\author[LANL]{A. R. Bishop},
\author[Gr,Er]{M. Hohenadler},
\author[Gw]{G. Schubert},\,
\author[Gw]{H. Fehske}
\address[Er]{Computing Center, University Erlangen, 91058 Erlangen, Germany}
\address[LANL]{Theoretical Division and Center for Nonlinear Studies, 
  Los Alamos National Laboratory, Los Alamos, NM 87545, USA}
\address[Gr]{Institute for Theoretical and Computational Physics, TU Graz, 8010 Graz, Austria}
\address[Gw]{Institute for Physics, Ernst-Moritz-Arndt University Greifswald, 17487 Greifswald, Germany}

\corauth[Name1]{Corresponding author. Tel: +49 9131 852 8737}

\begin{abstract}
  Exact results for the density of states and the ac conductivity of the
  spinless Holstein model at finite carrier density are obtained combining
  Lanczos and kernel polynomial methods.
\end{abstract}
\begin{keyword}
electron-phonon interaction, optical conductivity, Holstein polarons
% keywords here, in the form: keyword \sep keyword
% PACS codes here, in the form: 
\PACS 71.27.+a, 63.20.Kr, 71.10.-w, 71.38.-k
\end{keyword}
\end{frontmatter}
Optical measurements have proved the importance of electron-phonon (EP)
coupling and even polaron effects in several important classes of materials,
including one-dimensional (1d) MX chains, quasi-2d cuprate superconductors,
and 3d colossal magnetoresistance manganites. In all these materials, a
noticeable density of (polaronic) charge carriers is observed, which puts the
applicability of single-polaron theories into question, particularly in the
region of intermediate EP coupling strength and phonon frequency.

Recently, the photoemission spectra of many-polaron systems have been
investigated in the framework of the 1d spinless Holstein model,
\begin{equation}\label{homo}
  H\!
  =\!
  -t \sum_{\langle i,j\rangle} c^\dag_i c^{\nag}_j
  \!+\!\omega_0\sum_i b^\dag_i b^{\nag}_i
  \!-\!g\omega_0 \sum_i 
\hat{n}_i (b^\dag_i + b^{\nag}_i),
\end{equation}
describing tight-binding ($t$) electrons coupled locally ($g$) to
Einstein phonons ($\omega_0$), where $c^\dag_i$ ($b^\dag_i$)
denote the corresponding fermionic (bosonic) creation operators, and
$\hat{n}_i=c^\dag_i c^{\nag}_i$. Most notably, provided that the EP coupling
is not too strong, a density driven crossover from large polarons to weakly
dressed electrons has been found to occur \cite{HNLWLF04}. In the meantime,
this result was corroborated by cluster perturbation theory
(CPT) \cite{HWAF05}.

Here, we use Lanczos diagonalization and kernel polynomial expansion
methods \cite{WWAF05} to study the (linear) optical response of a
many-polaron system to an external (longitudinal) electric field, $ {\rm
  Re}\,\sigma(\omega) = D \delta(\omega)+ \sigma^{\rm reg}(\omega)$, the
regular part of which is given by
\begin{equation}\label{si}
\sigma^{\rm reg}(\omega)
=
\frac{\pi}{N} \sum_{n>0} 
\frac{|\langle n|\hat{\jmath}|m\rangle|^2}{\omega_{mn}}\, 
   \delta(\omega - \omega_{mn})\,,
\end{equation}
where $\omega_{mn}= E_m - E_n$, $E_n$ ($|n\rangle$) are the eigenvalues
(eigenstates) of our $N$-site coupled EP system with at most $M$ phonons, and
$\hat{\jmath}={\rm i}\, e t \sum_i (c^\dag_i c^{\nag}_{i+1} -
c^{\dag}_{i+1}c^{\nag}_i)$.  In addition, we calculate the partial densities
of states (DOS) $\rho^+(\omega)$ and $\rho^-(\omega)$, which are obtained
from the $k$-integrated single-particle spectral functions
\begin{eqnarray}\label{ipe}
A_{k}^{\pm}(\omega)
&=& 
\sum_{m} 
|\langle m^{(N_{el}\pm1)}|c_{k}^{\pm} |0^{(N_{el})}\rangle|^2
\nonumber \\
&&\qquad
\times
\delta [\,\omega\mp(E_m^{(N_{el}\pm1)}-E_0^{(N_{el})})]\,, 
\end{eqnarray}
$c_{k}^{+}=c_{k}^{\dagger}$, $c_{k}^{-}=c_{k}^{}$, where $A_{k}^{-}(\omega)$
[$A_{k}^{+}(\omega)$] is related to the [inverse (I)] photoemission (PE) of
an electron.

\begin{figure*}[th!]
    \centering
    \includegraphics[width=0.85\textwidth]{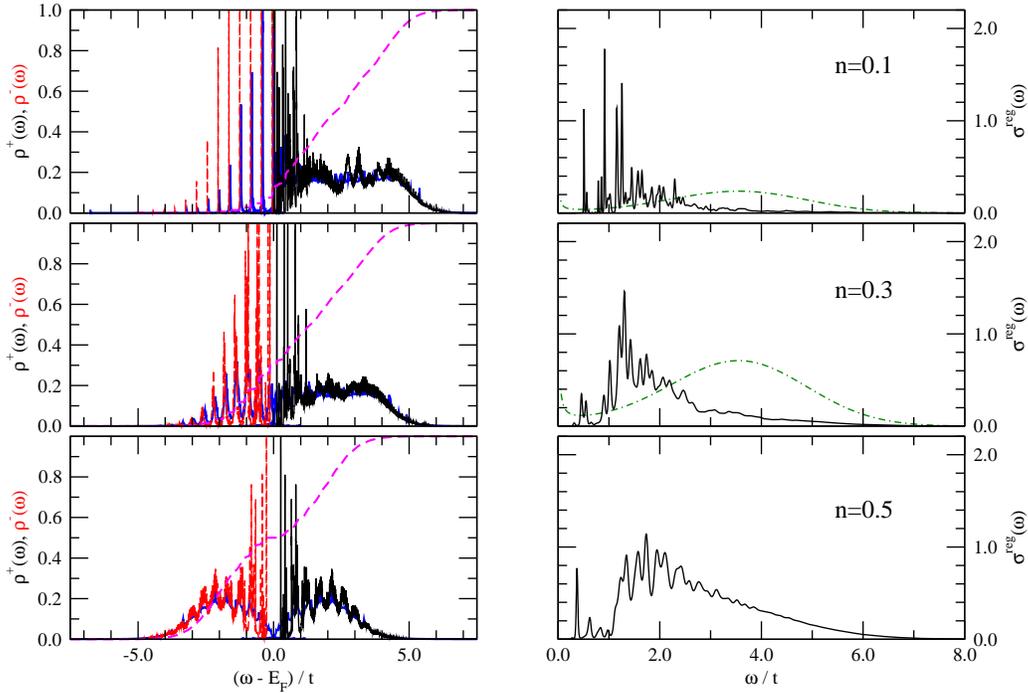}
    \caption{\label{fig1}
      (Color online)
      Left panel:
      Single-particle DOS $\rho^{-}(\omega)$ 
      (red dashed lines) and $\rho^{+}(\omega)$ (black solid lines) for
      $g^2=5$, $\omega_0/t=0.4$, and particle
      densities $n=0.1$ (one electron), $n=0.3$, and $n=0.5$ (half filling).
      Results are for a ten-site system with periodic boundary 
      conditions and $M\leq15$ dynamical phonons. The homogeneous
      $q=0$ lattice displacement was treated separately \cite{FHW00}, leading
      to truncation errors $<10^{-4}$. 
      Dashed (magenta) lines give the integrated DOS. CPT data for
      $\rho^{\pm}(\omega)$ included for comparison (blue lines) were partly
      extracted from \cite{HWAF05}.   
      Right panel: 
      Regular part of the optical conductivity $\sigma^{\rm reg}(\omega)$. 
      (Green) dashed-dotted lines: analytical strong-coupling result
      $\sigma^{\rm reg} (\omega) = \sigma_0\,n\, (\omega_0g)^{-1}\,
      \omega^{-1}\, \exp [(\omega-2g^2\omega_0)/(2g\omega_0)]^{2}$
      ($\sigma_0=8$) \cite{Emi93}.
    }
\end{figure*}

Figure~\ref{fig1} displays selected numerical results for the optical
properties of the 1d model~(\ref{homo}) for $g^2=5$, $\omega_0/t=0.4$
and various characteristic particle densities.

Starting with a {\it single electron} ($n=0.1$), we notice from the DOS that
there is a polaron feature at the Fermi level $E_{\rm F}$ (cf. the jump in
the integrated DOS at $E_{\rm F}$ and the negligible spectral weight of
$\rho^-(\omega<E_{\rm F})$). However, in view of the intermediate EP coupling
strength chosen, $\lambda=g^2\omega_0/2t=1$, the polaron is rather extended
(large).  Consequently, the polaron band is not far separated from incoherent
excitations, and $\sigma^{\rm reg}(\omega)$ strongly deviates from the
analytical strong-coupling result.  In particular, the maximum in
$\sigma^{\rm reg}$ occurs well below the small polaron value $4\lambda$.

At {\it finite carrier density}, the system shows diffusive transport. The
polarons are dissociated and the remaining electronic quasiparticles are
scattered by virtual phonons. The peaks in the PE part of the spectrum, which
for $n=0.1$ had reflected the Poisson-like distribution of phonons in the
ground state, now broaden significantly, and ultimately merge with the IPE
part into a wide incoherent band (see also the continuous increase of the
integrated DOS) \cite{HNLWLF04}.
 
In the {\it half-filled band case}, the model has a symmetry-broken
insulating charge density wave ground state accompanied by a distortion of
the lattice. This is because the EP coupling exceeds the critical interaction
strength for the Peierls instability: $g>g_c(\omega_0/t=0.4)\simeq 1.9$
\cite{FHW00}. Accordingly, we find a gap feature in the DOS and a
clear optical absorption threshold. Note that the CPT does not reproduce
the very small gap in the DOS, as it ``interpolates'' between the wavevectors
of the finite cluster.

In conclusion, increasing the particle density in the 1d spinless Holstein
model at intermediate EP coupling strengths, we observe a crossover from a
polaronic system to a metal composed of weakly dressed electronic
quasi-particles and finally to a Peierls insulator at half filling. These
transitions are reflected in significant changes of the optical spectra.

This work was supported by DFG under SPP1073,  KONWIHR Bavaria, 
NIC J\"{u}lich, and the US DOE.

\end{document}